\documentclass[aps, prx, twocolumn, amssymb, amsmath, superscriptaddress]{revtex4-1}

\usepackage{bm}
\usepackage{times}
\usepackage{graphicx}
\usepackage[colorlinks,citecolor=blue,linkcolor=blue]{hyperref}
\usepackage{array}
\usepackage{float}
\usepackage{colortbl}
\usepackage{multirow}
\usepackage{tabularx}

\begin{document}

\title{The direct and indirect optical absorptions of cubic BAs and BSb}

\author{Yanfeng Ge}
\affiliation{State Key Laboratory of Metastable Materials Science and Technology \& Key Laboratory for Microstructural Material Physics of Hebei Province,
School of Science, Yanshan University, Qinhuangdao, 066004, China}

\author{Wenhui Wan}
\affiliation{State Key Laboratory of Metastable Materials Science and Technology \& Key Laboratory for Microstructural Material Physics of Hebei Province,
School of Science, Yanshan University, Qinhuangdao, 066004, China}

\author{Xitong Guo}
\affiliation{State Key Laboratory of Metastable Materials Science and Technology \& Key Laboratory for Microstructural Material Physics of Hebei Province,
School of Science, Yanshan University, Qinhuangdao, 066004, China}

\author{Yong Liu}\email{yongliu@ysu.edu.cn}
\affiliation{State Key Laboratory of Metastable Materials Science and Technology \& Key Laboratory for Microstructural Material Physics of Hebei Province,
School of Science, Yanshan University, Qinhuangdao, 066004, China}
\date{\today}

\begin{abstract}
Recently, boron arsenide (BAs) has been measured high thermal conductivity in the experiments, great encouraging for the low-power photoelectric devices. Therefore, in the present work, we have systematically investigated the direct and indirect optical absorptions of BAs and BSb and the doping effect of congeners by using first-principles calculations. We obtain the absorption onset corresponding to the value of indirect bandgap by considering the phonon-assisted second-order optical absorptions. And the redshift of absorption onset, enhancement and smoothness of optical absorptions spectra are also captured in the temperature-dependent calculations. In order to introduce one-order absorptions into the visible range, the doping effect of congeners on optical absorptions is studied without the assists of phonon. It is found that the decrease of local direct bandgap after doping derives from either the small bandgap in the prototypical $\rm \uppercase\expandafter{\romannumeral3}$-$\rm \uppercase\expandafter{\romannumeral5}$ semiconductors or CBM locating at R$_c$ point. Thus, doping of congeners can improve the direct optical absorptions in visible range.

\end{abstract}

\maketitle

\section{Introduction}

The photoelectric devices have become the indispensable micro devices in the modern industrial society. As the future development trend of the lower powerful photoelectric devices, heat dissipation is the mast main factor restricting the performance of devices. Thus the versatile materials with high thermal conductivity ($\kappa$) become increasingly important. Carbon based materials have long been recognized as having higher thermal conductivity than other bulk materials. Particularly diamond and graphite have a record of $\sim$ 2000 W/(m$\cdot$K) at room temperature~\cite{Slack1972,Wei1993,Olson1993}, but their applications are seriously limited by the high cost and anisotropy, respectively.

By using a predictive first-principles calculation with Boltzmann transport approach, it is found that boron arsenide (BAs) with zinc-blende face-centered cubic structure have a remarkable $\kappa$ over 2000 W/(m$\cdot$K) limited by three-phonon scattering~\cite{Lindsay2013,Broido2013}. However, the high-order anharmonicity reduces $\kappa$ to 1400 W/(m$\cdot$K)~\cite{Lindsay2008,Feng2017}. The strong covalent bonding and the large mass ratio between B and As atom give rise to the acoustic bunching as well as the large frequency gap between acoustic phonons and optic phonons, which effectively increase intrinsic thermal conductivity.
Subsequently, BAs is experimentally isolated by the chemical vapor transport method~\cite{Li2018,Kang2018,Tian2018}. And the experimental data of high thermal conductivity, $\sim$1300 W/(m$\cdot$K) at room temperature, agrees well with the theoretical predictions.

Unfortunately, BAs has a global indirect bandgap~\cite{Nwigboji2016} of $\sim$1.3 eV and large local direct bandgap of $\sim$ 3.4 eV, similar to silicon~\cite{Chu1974,Wang2012}. This obviously limits the one-order absorptions process of the visible range in the standard model calculations.
But the second-order absorption process~\cite{Giustino2017,Zacharias2015,Noffsinger2012} helps silicon apply to solar cell with tolerable photoelectric conversion efficiency. It should involve the electron-phonon coupling and utilize the phonons to ensure the momentum conservation of electronic transition. When the photon energy is greater than the indirect bandgap minus (plus) the absorbed (emitted) phonon energy, the phonons can assist the indirect optical absorptions process. Besides silicon, phonon-assisted optical absorptions are also studied in other materials~\cite{Peelaers2015,Monserrat2018,Morris2018,Menendez2018,Karsai2018}, which all have indirect bandgap but application potential in photoelectric devices. Considering the powerful cooling performance of BAs and the comparability of electronic structure to silicon, BAs holds plenty of promise in the next-generation photoelectric devices. Thus it is essential to study the optical properties of BAs seriously, and a full description of optical absorption requires the calculation including electron-phonon coupling.
Moreover, the executable methods are also necessary to improve the absorptions process in visible range.
Some common methods, such as pressure~\cite{Ackland2001,Boudjemline2011} and doping~\cite{Chimot2005,Lindsay2007,Guemou2012}, can influence the electronic structure and change the physical properties, so may also improve the one-order absorption process in the visible range.

In this paper, based on the first-principles calculations, we investigate the one-order and second-order optical absorptions of BAs and BSb, which are both predicted to have high thermal conductivity in the previous works. The results show that the main optical absorption peaks appear in the ultraviolet region around 7.0 eV. The absorption onset is in coordination with the indirect bandgap, and also red shifts with the increase of temperature. The absorption spectra in the visible range of BAs and BSb are similar to that observed in other indirect bandgap semiconductors~\cite{Noffsinger2012}. And the doping of congeners can enhance the one-order optical absorptions in visible range, due to the reduction of local direct bandgap.

\section{Methods}

Technical details of the calculations are as follows. All calculations in this work were carried out in the framework of density functional theory (DFT) with General gradient approximation (GGA) in the Perdew-Burke-Ernzerhof (PBE) implementation~\cite{Perdew1996}, as implemented in the QUANTUM ESPRESSO~\cite{Giannozzi2009}. The ion and electron interactions are treated with the norm-conserving pseudopotentials~\cite{Troullier1991}.
The hybrid Hartree-Fock+DFT functional of HSE was used in order to obtain the accurate bandgap and the direct optical absorption in the standard model calculations. By requiring convergence of results, the kinetic energy cutoff of $600$~eV and the Monkhorst-Pack $k$-mesh of 30$\times$30$\times$30 (20$\times$20$\times$20) were used in all calculations about the ground-state properties of primitive cell (cubic crystal cell). The phonon spectra and electron-phonon coupling were calculated on a 30$\times$30$\times$30 $q$-grid using the density functional perturbation theory (DFPT)~\cite{Baroni2001} and maximally localized wannier functions~\cite{Marzari2012,Giustino2007,Noffsinger2010,Ponce2016}.
Phonon-assisted optical absorption process could be analyzed by the second-order time-dependent perturbation theory with electron-phonon coupling~\cite{Giustino2017,Zacharias2015,Noffsinger2012,Peelaers2015}, which was applied to some semiconductors, such as silicon, SnO and so on.
The bandgap of HSE functional also gave the value of the scissor shift of the gap in the calculations about phonon-assisted indirect absorption process.

\section{Results}

\begin{figure}[ht!]
\centerline{\includegraphics[width=0.4\textwidth]{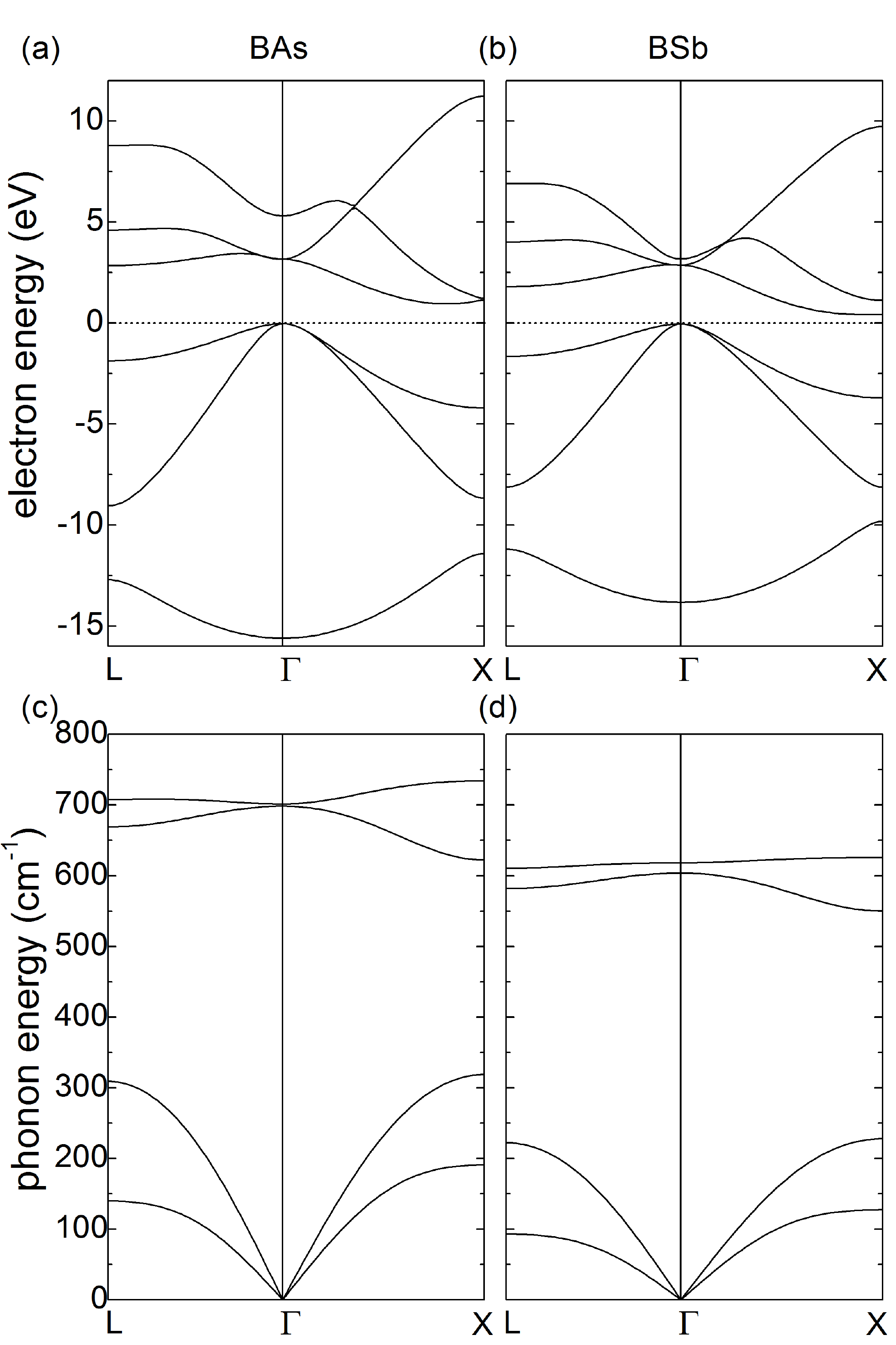}}
\caption{{\bf Band structure and phonon spectra.} (a) Band structures of BAs and (b) BSb with primitive cell calculated using PBE functional. Both materials have VBM at $\Gamma$ point, but the CBM of BAs locates on $\Gamma$-X (0.5,0.0,0.5) line, different from the X point in BSb. (c) Phonon spectra of BAs and (b) BSb. Two transverse modes are degenerate alone $\Gamma$-L (0.5,0.5,0.5) and $\Gamma$-X lines, thus phonon spectra show four branches.
\label{fig:1band}}
\end{figure}

The lattice structure of BAs and BSb is a face-centered cubic structure with space group of F$\bar{4}$3m, where B atoms locate the tetrahedral center of the four nearest neighbor V-group atoms with the B-V-B bond angle of 109.47$^\circ$. The lattice constants of BAs and BSb are 4.80 ${\rm \AA}$ and 5.30 ${\rm \AA}$, respectively, agreement well with the previous works~\cite{Hart2000,Ahmed2007,Cui2009,Deligoz2007}. They are both indirect bandgap semiconductors and have silicon-like band structures. As shown in Figs.~\ref{fig:1band}(a) and (b), the valence band maximum (VBM) locates at $\Gamma$ point and conduction band minimum (CBM) locates at $\Gamma$-X line (BAs) or X point (BSb). The bandgap of BAs and BSb in the band structures calculated using PBE functional are 1.31 and 0.80 eV, respectively. In order to correct the underestimated effect of PBE functional, the hybrid functional of HSE are used to obtain the accurate bandgap, which increases to 1.58 (1.06) eV in BAs (BSb). The primitive cell of BAs or BSb includes two atoms thus the phonon spectra should have six phonon branches (three acoustic and three optical branches). But the symmetries alone $\Gamma$-L and $\Gamma$-X lines lead to the double degenerate transverse acoustic modes and transverse optical modes, as shown in Figs.~\ref{fig:1band}(c) and (d). The heavy Sb atom also brings about the lower frequency in the whole spectrum of BSb than that of BAs. It worth noting that the LO-TO splitting of BSb (14.3 cm$^{-1}$) is significantly greater than that of BAs (3.1 cm$^{-1}$), because of the bigger difference of ionicity between B and Sb elements.

\begin{figure}[ht!]
\centerline{\includegraphics[width=0.5\textwidth]{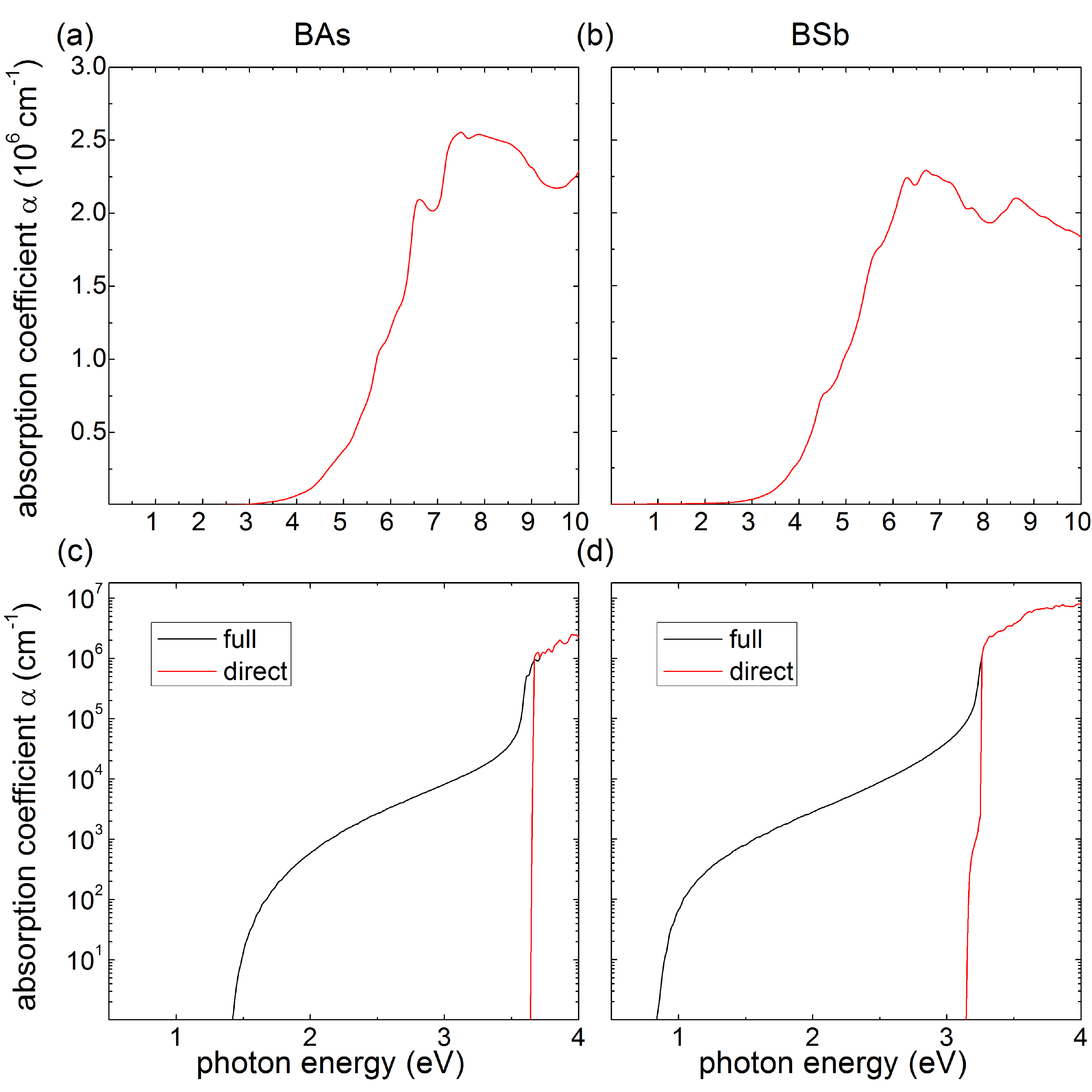}}
\caption{{\bf Full optical absorption spectra.} (a) The direct optical absorption spectra of BAs and (b) BSb in the standard model calculations with HSE functional. (c) The full optical absorption spectra of BAs and (d) BSb with the phonon-assisted indirect absorption process in the low-energy region (logarithmic scale). It is observed that the absorption onset transfer from the local direct bandgap to the global indirect bandgap.
\label{fig:2onset}}
\end{figure}

\subsection{Phonon-assisted optical absorptions}

In the standard model calculations including one-order absorption, the direct optical absorption spectra can be obtained by the dielectric function as $\alpha(\omega)=\sqrt{2}\omega[\sqrt{\varepsilon_1(\omega)^2+
\varepsilon_2(\omega)^2}-\varepsilon_1(\omega)]^{1/2}$, where $\varepsilon_1(\omega)$ and $\varepsilon_1(\omega)$ are the real and imaginary parts of frequency-dependent complex dielectric function.
As shown in Figs.~\ref{fig:2onset}(a) and (b), the direct optical absorption increases sharply between 4.5 and 6.5 eV, and the main peaks appear in the ultraviolet region around 7.0 eV. There is almost not optical absorption in the visible range (1.62$\sim$3.11 eV), limited by the large local direct bandgap in BAs and BSb. The direct absorption onset occurs at the minimum direct bandgap [Figs.~\ref{fig:2onset}(c) and (d)], corresponding to the transition between valence state (bonding $p$ states: $\Gamma_{15v}$) and conduction state (antibonding $p$ states: $\Gamma_{15c}$) at $\Gamma$ point~\cite{Hart2000}.
After considering the phonon-assisted indirect optical absorption at the temperature of 100 K, the redshift of absorption onset is shown obviously by the black solid line in Figs.~\ref{fig:2onset}(c) and (d). And the deviate of indirect absorption onset and the value of indirect bandgap is attributed to the smearing treatment of $\delta$ function about energy conservation in the Fermi's golden rule expression of phonon-assisted absorption coefficient~\cite{Marzari2012}.
The optical absorption spectra of BAs and BSb in the visible range ($<$ local direct bandgap) are similar to the changing curve of silicon in the previous work~\cite{Marzari2012}. This results indicate that the primary BAs and BSb can also be applied to the photoelectric devices even though the existence of indirect bandgap.

\begin{figure}[ht!]
\centerline{\includegraphics[width=0.5\textwidth]{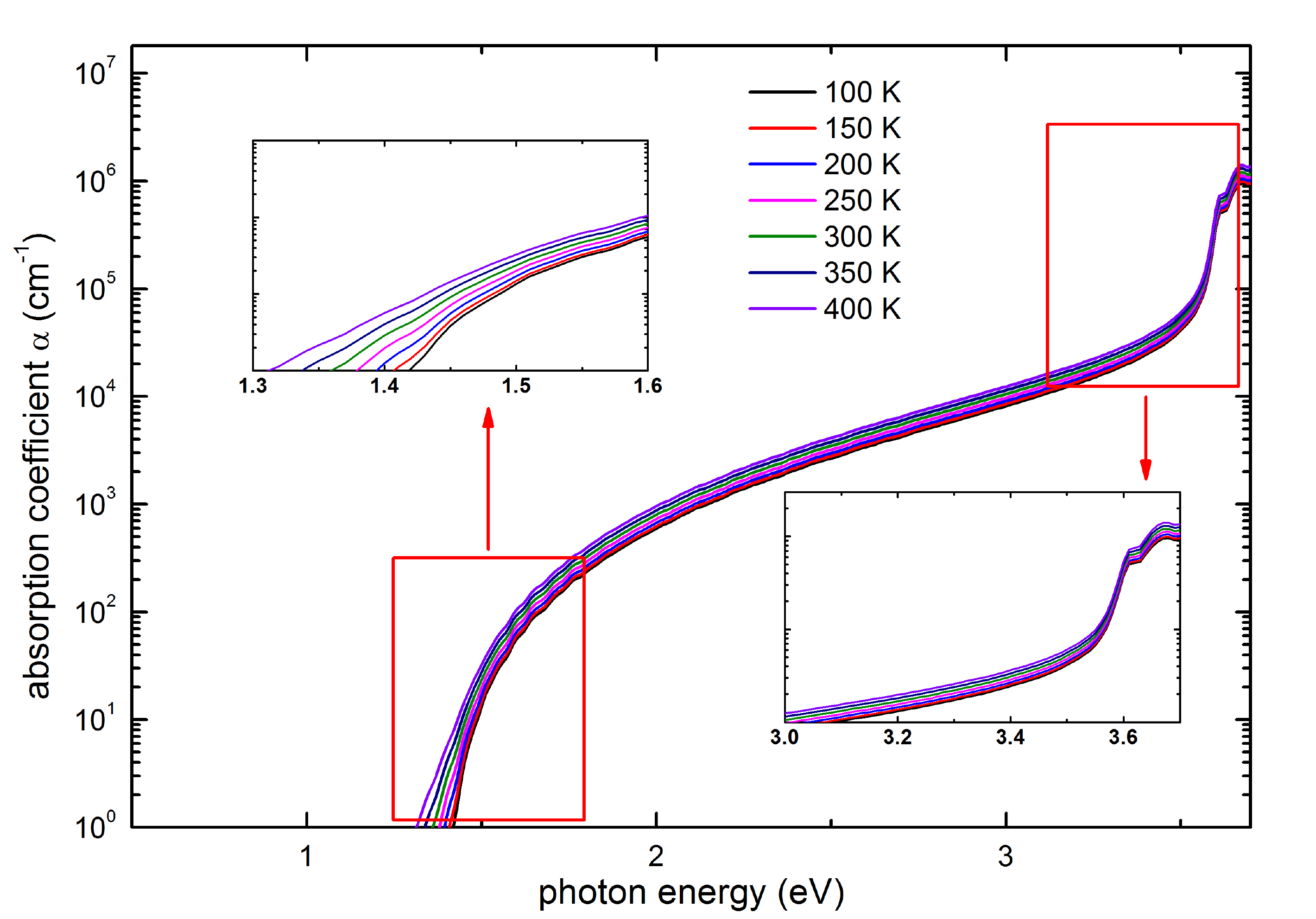}}
\caption{{\bf Temperature-dependent optical absorption spectra.} The full optical absorption spectra of BAs in low-energy region under different temperatures. The inset details are the optical absorption spectra around the values of indirect and direct bandgap.
\label{fig:3temp}}
\end{figure}

Fig.~\ref{fig:3temp} shows the temperature-dependent optical absorption spectra of BAs. Note that the temperature effects on electronic state energies, including the lattice expansion and electron-phonon renormalization~\cite{Giustino2010,Gonze2011,Antonius2014,Monserrat2016}, are not considered in our calculations. But the temperature-dependent results still capture two features of temperature effects. One is the redshift of $\sim$0.1 eV from 100 K to 400 K, similar to the universal phenomenon in other materials~\cite{Peelaers2015,Monserrat2018,Morris2018,Menendez2018}. But, the incomplete temperature effects in our calculations result in the small redshift of indirect optical absorption onset. And the other one is the enhancement and smoothness of optical absorption spectra at high temperature. With the increase of temperature, there are more phonons to assist the optical absorptions, and the attainable range of final state from the same initial state is broadened and the absorption cross section increases, thus the temperature gives rise to above effects, which are also present in BSb.

\subsection{Doping effect of congeners}

\begin{figure}[h]
\centerline{\includegraphics[width=0.5\textwidth]{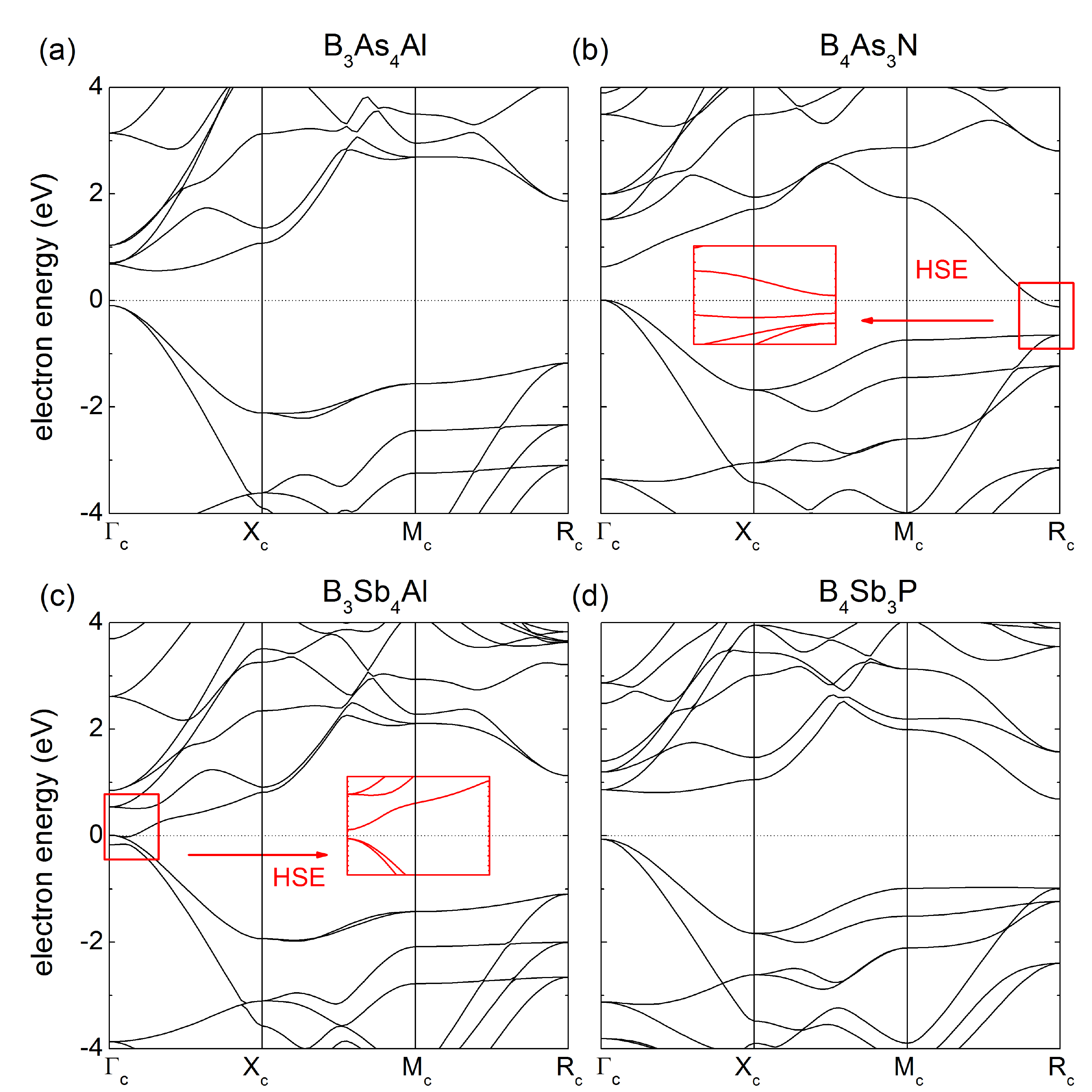}}
\caption{{\bf Doping effect of congeners on the band structures.} (a) The band structures of B$_3$As$_4$Al, (b) B$_4$As$_3$N, (c) B$_3$Sb$_4$Al and (d) B$_4$Sb$_3$P with cubic crystal cell calculated using PBE functional. The inset details in (b) and (c) shows that B$_4$As$_3$N and B$_3$Sb$_4$Al transforms into semiconductor in the HSE functional. The corner mark of symmetry points signifies that the K points are in the reciprocal space of cubic crystal cell.
\label{fig:4band}}
\end{figure}

Although the phonons assist the indirect optical absorption in the visible range, the second-order process is weaker than one-order process and has small correction in the higher energy region ( $>$ local direct bandgap), where one-order absorption is the dominating process. So, except for the phonon-assisted indirect absorption process, we also explore the way to improve the one-order absorption process in visible range. Consider that the doping of congeners has been successfully used to improve the performance of $\rm \uppercase\expandafter{\romannumeral3}$-$\rm \uppercase\expandafter{\romannumeral5}$ semiconductor~\cite{Chimot2005,Lindsay2007,Guemou2012}, we systematically investigate the doping effect of congeners on the optical absorption. And the calculations about doping effect do not refer the electron-phonon coupling.

\begin{table}[htp!]
\caption{Bandgap of B$_4$As$_4$ and B$_4$Sb$_4$ with the doping of congeners calculated using HSE functionals.}
\begin{tabular*}{6cm}{@{\extracolsep{\fill}}ccccccccc}
\hline\hline
        &   B$_3$As$_4$X (eV) &   & B$_3$Sb$_4$X (eV) \\
\hline   Al &  1.13 & Al   & 0.18   \\
         Ga &  0.63 & Ga   & metal  \\
         In &  0.91 & In   & metal  \\
\hline
        &   B$_4$As$_3$X (eV) &   & B$_4$Sb$_3$X (eV) \\
\hline   Sb &  1.39 & As   & 1.13   \\
         N  &  0.36 & N    & metal  \\
         P  &  1.72 & P    & 1.20   \\
\hline \hline
\end{tabular*}
\label{tab:bandgap}
\end{table}

\begin{figure*}[t!]
\centerline{\includegraphics[width=0.9\textwidth]{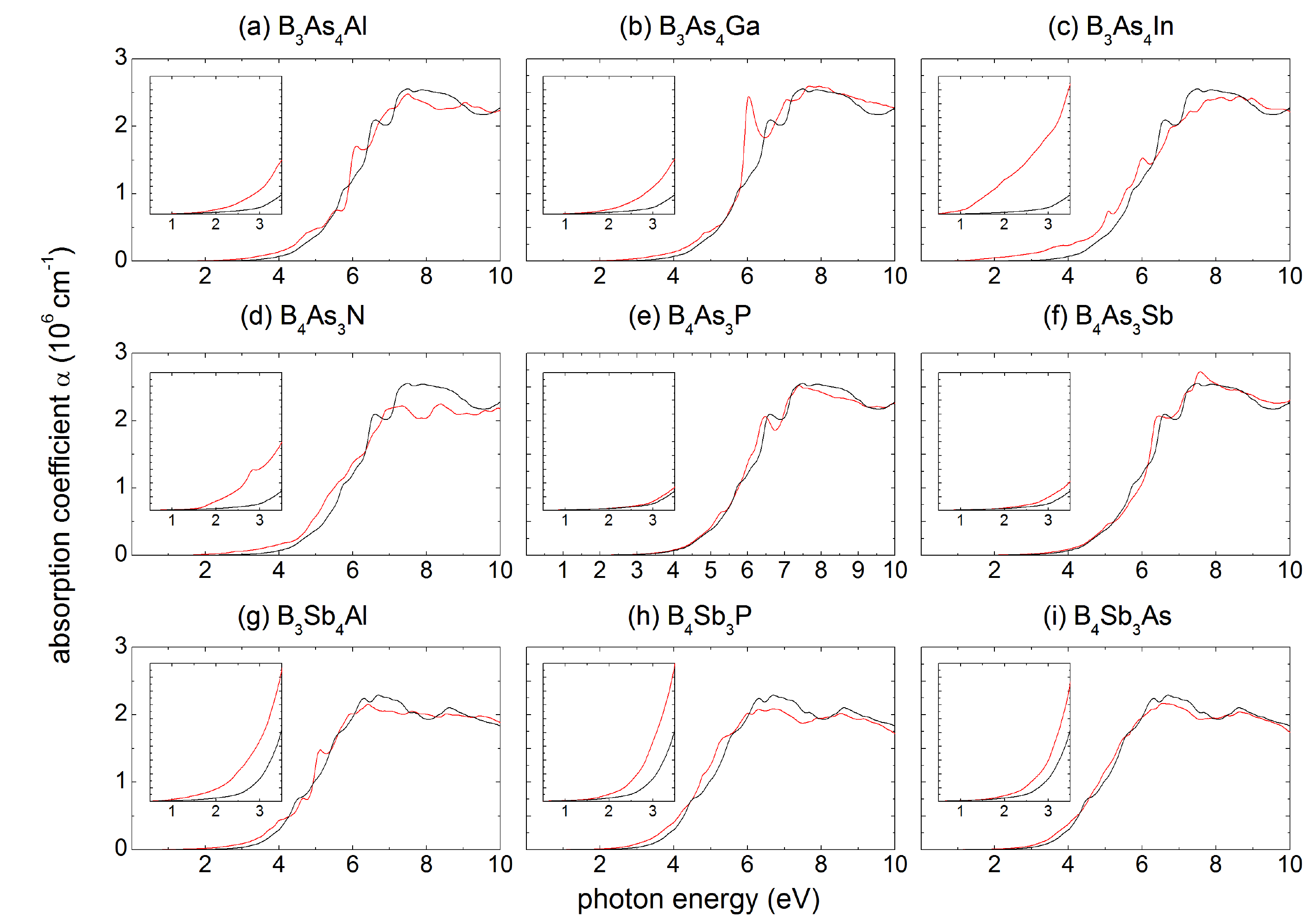}}
\caption{{\bf Doping effect of congeners on the optical absorption spectra.} (a) The direct optical absorption spectra of B$_3$As$_4$Al, (b) B$_3$As$_4$Ga, (c) B$_3$As$_4$In, (d) B$_4$As$_3$N, (e) B$_4$As$_3$P, (f) B$_4$As$_3$Sb, (g) B$_3$Sb$_4$Al,(h) B$_4$Sb$_3$P and (i) B$_4$Sb$_3$As in the standard model calculations. The black line is the optical absorption spectra of BAs or BSb. The inset details are zoom-in view of absorption spectra in low-energy region (0.5$\sim$3.5 eV)
\label{fig:5absor}}
\end{figure*}

In order to study the doping effect, we adopt the face-centered cubic (FCC) crystal (4 times primitive cell) as the periodic unit. For B$_4$As$_4$ and B$_4$Sb$_4$, one atom (B, As, and Sb) is substituted by one atom of same main group, such as B$_3$As$_4$Al and B$_4$As$_3$N. In the calculations of electronic structures, it is found that BAs-based systems are all the semiconductors, as summarized in Tab.~\ref{tab:bandgap}. However, there are only three semiconducting BSb-based systems. The band structures after doping are plotted in Fig.~\ref{fig:4band}. It is obvious that the four systems have a much smaller 'direct' bandgap than that of prototypical structure seemingly. But, what needs to pay attention is that the X point (0.5, 0.0, 0.5) in the reciprocal space of primitive cell is folded at $\Gamma_c$ point (0.0, 0.0, 0.0) in the reciprocal space of FCC crystal cell. Thus the folding leads to the pseudo direct bandgap and the doping of congeners hasn't really changed the feature of indirect bandgap in B$_4$As$_4$ or B$_4$Sb$_4$. Among all the doping systems, B$_3$Sb$_4$Al has the smallest bandgap in the HSE functional [Tab.~\ref{tab:bandgap}], but it is metallicity with the GGA calculations, which also appears in B$_4$As$_3$N [Fig.~\ref{fig:4band}].

Figure.~\ref{fig:5absor} plot the direct optical absorption spectra of all semiconducting doping systems.
By comparing the B$_3$As$_4$X (X=Al, Ga, In), as shown in upper panels in Fig.~\ref{fig:5absor}, it is observed that the enhancement of absorptions increases with the increase of atomic number of doping elements. We can analyze the reasons simply by the band structures of $\rm \uppercase\expandafter{\romannumeral3}$-As compounds~\cite{Vurgaftman2001,Ahmed2007,Singh2007}. With the increase of atomic number of $\rm \uppercase\expandafter{\romannumeral3}$-group elements,
$\rm \uppercase\expandafter{\romannumeral3}$-As compounds become from indirect bandgap (BAs) to direct bandgap at $\Gamma$ point (GaAs), even metal (InAs). It is palpable that B$_3$As$_4$X has smaller and smaller local direct bandgap at $\Gamma$ point in the reciprocal space of diatomic primitive cell, so doping can improve the optical absorption of B$_3$As$_4$X in visible range. However, B-$\rm \uppercase\expandafter{\romannumeral5}$ compounds are all indirect bandgap semiconductor thus has little doping effect on optical absorption, except for B$_4$As$_3$N, as shown in middle panels in Fig.~\ref{fig:5absor}. The calculation of B$_4$As$_3$N shows that the CBM locates at R$_c$ point (0.5,0.5,0.5), different from the $\Gamma_c$-X$_c$ line of B$_4$As$_3$P and B$_4$As$_3$Sb. So B$_4$As$_3$N has a smaller local direct bandgap and higher optical absorption in visible range.
The lower panels in Fig.~\ref{fig:5absor} show that all doping-BSb systems can increase the absorption in visible range. B$_3$Sb$_4$Al has small local direct bandgap at $\Gamma$ point, similar to B$_3$As$_4$X. Nevertheless, B$_4$Sb$_3$P and B$_4$Sb$_3$As have the CBM locating at R$_c$ point, similar to B$_4$As$_3$N.

\section{Discussion}

In the present work, we have systematically investigated the direct and indirect optical absorptions of BAs and BSb and the doping effect of congeners by using first-principles calculations. Due to the indirect bandgap and the large local direct bandgap of BAs and BSb, the one-order optical absorption in visible range is forbidden transition and the second-order absorption becomes the dominant process. Therefore, considering the phonon-assisted indirect optical absorption, we obtain the absorption onset corresponding to the value of indirect bandgap. Moreover, the redshift of absorption onset, enhancement and smoothness of optical absorption spectra are also captured in the temperature-dependent calculations. In view of the lower second-order absorption than one-order absorption and the importance of introducing one-order absorptions into the visible range, the doping effect of congeners on optical absorptions is studied without the assist of phonon. The main result is that the small local direct bandgap can improve the one-order optical absorption in visible range with two different reasons. One is that the small bandgap in the prototypical $\rm \uppercase\expandafter{\romannumeral3}$-$\rm \uppercase\expandafter{\romannumeral5}$ semiconductor including doping atom lead to the small local direct bandgap of doping systems. The other one is that CBM locating at R$_c$ point also generate small local direct bandgap.

In the future, inclusion of full-scale temperature effects, lattice expansion and electron-phonon renormalization, may improve the predictions of temperature-dependent optical absorptions and thus better guide the experiments. Furthermore, it's also important to notice that electron-hole Coulomb interaction is not taken into account, since the bandgap of HSE calculations is in good agreement with the experiments about $\rm \uppercase\expandafter{\romannumeral3}$-$\rm \uppercase\expandafter{\romannumeral5}$ semiconductors~\cite{Heyd2005,Mosesa2011,Nicklasa2010}. In fact, the band edge wave functions in the indirect bandgap semiconductors locate at different K points in reciprocal space, leading to the small overlap of wave functions, as well as the small Coulomb interaction between electrons and holes. Therefore, the combination of phonon-assisted spectra and HSE calculations can simulate the optical properties well, without the need to account for excitonic effects.

\begin{acknowledgments}
This work was supported by the NSFC (Grants No.11747054), the Specialized Research Fund for the Doctoral Program of Higher Education of China (Grant No.2018M631760), the Project of Heibei Educational Department, China (No. ZD2018015 and QN2018012), and the Advanced Postdoctoral Programs of Hebei Province (No.B2017003004).
\end{acknowledgments}

\end{document}